\documentclass[notitlepage]{article}%
\usepackage{amsmath}
\usepackage{amsfonts}
\usepackage{amssymb}
\usepackage{graphicx}%
\providecommand{\U}[1]{\protect\rule{.1in}{.1in}}

\begin{document}

\title{ Hints of New Physics from $D_{s}^{+}\rightarrow K^{+}\upsilon\overline
{\upsilon},$ $\ ~D^{0}\rightarrow\pi^{0}\upsilon\overline{\upsilon}%
~$and$~D_{s}^{+}\rightarrow D^{+}\upsilon\overline{\upsilon}~$ $~ $Decays}
\author{Shakeel Mahmood$^{{\footnotesize (1)}}$, Farida Tahir, Azeem Mir\\\textit{Comsats Institute of Information Technology,}\\\textit{\ Department of Physics, Park Road,\ Chek Shazad,Islamabad}\\$^{({\footnotesize 1)}}${\footnotesize shakeel\_mahmood@hotmail.com}}
\date{}
\maketitle

\begin{abstract}
We study rare decays $D_{s}^{+}\rightarrow K^{+}\upsilon\overline{\upsilon
},~D_{s}^{+}\rightarrow D^{+}\upsilon\overline{\upsilon}~$and$~D^{0}%
\rightarrow\pi^{0}\upsilon\overline{\upsilon}~$in NSIs. We calculate the NSIs
Branching ratios of these decays. There is a strong dependence of these on new
physics parameter. They provide, stringent constraints on $\epsilon_{\tau\tau
}^{uL}$, $\epsilon_{\tau\tau}^{dL}~$and $\epsilon_{\alpha\beta}^{dL}~$%
($\alpha,\beta=e,\mu$).

%

PACS numbers:
12.60.-i, 13.15.+g, 13.20.-v

\end{abstract}

\section{Introduction}

Standard Model (SM) of particle physics is a highly successful model. It has
been tested to high degree of precision up to 1\%. The 2012 discovery of higgs
\cite{higgs dis}\cite{higgs dis1}\cite{higgs dis2} and confirmation in 2014
\cite{SM higgs} that it is SM higgs further strengthen the model. But, despite
of all these successes it is not the end of book. There are many questions
which are not tackled by SM yet. It is silent about the possible pattern for
particles masses, known as mass hierarchy problem. Mass of top can not be
predicted without experimental evidence. It has no answer to the generations
of leptons and quarks. Neutrino mass is a concrete experimental evidence
against standard model. It is generally believed that SM is a low energy
approximation of some more fundamental theory. Experiments provide us the
precision test of SM as well as the possibility of new physics (NP) yet
unknown. Presently, one way of exploring NP is at very high energy machines
like LHC. Other way of search is at B-factories and BES-III which are although
operating at \ very low energies as compared to LHC but due to high
intensities have low background. So they can also be used for search of NP. NP
can be found in the leptonic and semileptonic decays of the mesons. We have
three types of decays in which it can be searched

(1) Flavor Changing Neutral Currents \ (FCNC), highly suppressed in SM

(2) Lepton Flavour Violating decays (LFV), not allowed in SM

(3) Lepton Number Violating decays (LNV), not allowed in SM

Rare decays occur at loop level and involve FCNC. These decays can be
represented by
\[
M\longrightarrow M^{^{\prime}}l_{\alpha}l_{\beta}%
\]
where $M=K,D,B$ mesons and $M^{^{\prime}}=\pi,K\,,D$ mesons and $M>M^{^{\prime
}}$and $l\ $is a charge lepton or a neutrino. At quark level it can be written
as
\[
q\longrightarrow q^{^{\prime}}l_{\alpha}l_{\beta}%
\]
which is represented in SM by%
\[
L_{eff}^{SM}=-2\sqrt{2}G_{F}(\overline{\nu}_{\alpha}\gamma_{\mu}L\nu_{\alpha
})(\overline{f}\gamma^{\mu}Pf)
\]

where $\alpha$ corresponds to the light neutrino flavor, $f$\ \ denote a
charged lepton or quark, where we are only dealing with quarks and $P$=
$R$\ or $L$\ with $R(L)$\ $=(\frac{1\pm\gamma_{5}}{2})$.

(2) $\alpha\neq\beta$; strictly forbidden due to lepton flavor violation, so
having no contribution in SM, only possible in Non standard interactions
(NSIs) \cite{S. Davidsona}
\[
L_{eff}^{NSI}=-2\sqrt{2}G_{F}\left[  \underset{\alpha=\beta}{\sum}%
\epsilon_{\alpha\beta}^{fP}(\overline{\nu}_{\alpha}\gamma_{\mu}L\nu_{\beta
})(\overline{f}\gamma^{\mu}Pf)+\underset{\alpha\neq\beta}{\sum}\epsilon
_{\alpha\beta}^{fP}(\overline{\nu}_{\alpha}\gamma_{\mu}L\nu_{\beta}%
)(\overline{f}\gamma^{\mu}Pf)\right]
\]
Here $\epsilon_{\alpha\beta}^{fP}$ is the parameter for NSIs, which carries
information about dynamics. NSIs are considered to be well-matched with the
oscillation effects along with new features\ in neutrino searches \cite{P.
Huber}\cite{P. Huber1}\cite{P. Huber2}\cite{P. Huber3}\cite{P. Huber4}\cite{P.
Huber5}\cite{P. Huber6}. NSIs may conserve flavor $\alpha=\beta,$for this we
have $\epsilon_{ee}^{fP},$ $\epsilon_{\mu\mu}^{fP}$ and $\epsilon_{\tau\tau
}^{fP}$ known as flavour diagonal ($FD$). It can violate flavor conservation
$\alpha\neq$ $\beta,$ for which we have $\epsilon_{e\mu}^{fP},$ $\epsilon
_{e\tau}^{fP},$ $\epsilon_{\mu e}^{fP},$ $\epsilon_{\mu\tau}^{fP},$
$\epsilon_{\tau e}^{fP},$ and $\epsilon_{\tau\mu}^{fP}$ known as Flavor non
diagonal ($FND$). Constraints on NSI parameter $\epsilon_{\alpha\beta}^{fP}$
have been studied in References \cite{C. H. Chen}\cite{Davidson}\cite{V. D.
Barger}\cite{Z. Berezhiani}. From scattering in leptonic sectors $(f$ is
lepton$),$ constraints are determined for first two generations $\epsilon
_{ll}^{fP}$ ($l=e$,$\mu$ ) by tree level processes and could be limited at
$O(10^{-3})$ by future $\sin^{2}\theta_{W}$ experiments. For third generation
($\tau$) we study decays which occur at loop level. KamLAND data \cite{Phys.
Rev} and solar neutrino data \cite{K. Eguchi et a}\cite{J. Boger et al.} can
improve the third generation ($\tau$)\ \ limit to $(0.3)$ \cite{C. H. Chen}\ .
Although, the constraints on $\epsilon_{\tau l}^{fP}$ are given by the
precision experiments but they are bounded by $O(10^{-2})$ \cite{C. H.
Chen}\cite{M. L. Mangano et al}.

When decays are measured experimentally we actually observe contributions from
SM as well as beyond SM or NP. Thus, when we declare that there is no NP we
have to be careful that we have not absorbed such new evidence into SM
physics. There are following approaches for the search of new physics in decays

(1) We predict the decay rate of a single process with known couplings and
compare it to experiments

(2) We make measurements of CKM parameters and compare them that they agree or not

(3) The same quantity is measure in several ways, even if it cannot be
predicted by SM e.g. CP violation.

The bottom line drown from above mentioned approaches is that we have NP in
the rare decays of mesons. Semileptonic decays of $K$ and $B$ mesons have and
will continue their role for exploring NP. But for $D$ sector due to smallness
of the branching ratios in SM and lack of experimental data, semileptonic
charm physics is difficult to study. But now data from BES-III, B factory,
Super-B and LHC-b for the rare decays which will improve our knowledge of
charm physics. A theoretical estimate for CC (charge currents) decays
$D_{s}^{+}\rightarrow D^{0}e^{+}\upsilon_{e},B_{s}^{0}\rightarrow B^{+}%
e^{-}\overline{\upsilon}_{e},D_{s}^{+}\rightarrow D^{+}e^{+}e^{-}$and
$B_{s}^{0}\rightarrow B^{0}e^{+}e^{-}$ is given in \cite{Hai-Bo} for future
data at different luminosities of these machines. Theoretical values of NSIs
could also be calculated for FCNC in charm decays. We select $D$ $(D_{s}%
^{+}\rightarrow K^{+}\upsilon\overline{\upsilon},~D^{+}\rightarrow
D^{0}\upsilon\overline{\upsilon},D^{0}\rightarrow\pi^{0}\upsilon
\overline{\upsilon})$ for this purpose and analysis them in the frame work of
NSIs. First of all we give standard model treatment in section 2. Section 3
and 4 present NSIs of $D_{s}^{+}\rightarrow K^{+}\upsilon\overline{\upsilon
},~D^{0}\rightarrow\pi^{0}\upsilon\overline{\upsilon}~$and $D_{s}%
^{+}\rightarrow D^{0}\upsilon\overline{\upsilon}$ respectively. Then we
summarize and discuss in section 5 and finally conclusion is provided in
section 6.

\section{ Rare Decays of D in The Standard Model}

$c\rightarrow u~\upsilon\overline{\upsilon}~$is a FCNC process for which SM
diagrams are shown in fig 1. Such processes proceed through the exchange of
the down type quarks quark in the loop contrary to $B$ and $K$ mesons, even
for $b$ quark $CKM$\ matrix element $V_{cb}$ is so small that overall effect
is negligible. The masses of $s$ and $d$ are smaller then the non-perturbative
QCD scale $\Lambda_{QCD}$, so \emph{GIM} cancellation is perfect for short
distance . It's SM Hamiltonian is
\[
H_{eff}^{SM}=\frac{G_{F}}{\sqrt{2}}\frac{\alpha_{em}}{2\pi\sin^{2}\theta_{W}%
}\underset{\alpha,\beta=e,\mu,\tau}{\Sigma}[V_{cs}^{\ast}V_{us}X(x_{s}%
)+V_{cb}^{\ast}V_{ub}X(x_{b})]\times(\overline{u}c)_{V-A}(\nu_{\alpha
}\overline{\nu}_{\beta})_{V-A}%
\]
but for $D_{s}^{+}\rightarrow K^{+}\upsilon\overline{\upsilon}$ the dominant
contribution comes from long distance. It is free from QCD complications
because they can be absorbed with tree level process $D_{s}^{+}\longrightarrow
K^{0}e^{+}\nu_{e}.~$Long distance SM branching ratio $4\times10^{-16}$of
$D_{s}^{+}\rightarrow K^{+}\upsilon\overline{\upsilon}~$is out of approach for
any existing detector.%

\ {}%
%

\ {}{}{}%

$\underset{\text{%
\  Figure 1.%
}~}{%
\begin{tabular}
[c]{|l|}\hline%
{\parbox[b]{2.872in}{\begin{center}
\includegraphics[
natheight=2.445700in,
natwidth=6.527600in,
height=1.049in,
width=2.872in
]%
{NEINK200.wmf}%
\\
$S\tan dard$ $Model$ $Diagrams$ $for$ $c\longrightarrow u\nu\overline
{\upsilon}$%
\end{center}}}
\\\hline
\end{tabular}
\ \ \ \ \ \ \ }$

\section{NSIs in $D_{s}^{+}\longrightarrow K^{+}\nu_{\alpha}\overline{\nu
}_{\beta}$}%

\begin{tabular}
[c]{l}%
$\underset{\text{%
\  Figure  2.%
}~\ }{%
{\parbox[b]{1.9796in}{\begin{center}
\includegraphics[
natheight=3.146200in,
natwidth=4.531600in,
height=1.6172in,
width=1.9796in
]%
{NEINK201.wmf}%
\\
$NSI$ $diagram$ $for$ $c\longrightarrow u\nu\overline{\upsilon}$
\end{center}}}
}$%
\end{tabular}
%

\ {}{}{}%

The NSIs diagram $c\longrightarrow u\nu_{\alpha}\overline{\nu}_{\beta}$ is
given in fig 2 represented by%
\[
H_{c\longrightarrow u\nu_{\alpha}\overline{\nu}_{\beta}}^{NSI}=\frac{G_{F}%
}{\sqrt{2}}(\frac{\alpha_{em}}{4\pi\sin^{2}\theta_{W}}V_{cd}V_{ud}^{\ast
}\epsilon_{\alpha\beta}^{dL}\ln\frac{\Lambda}{m_{W}})(\overline{\nu}_{\alpha
}\nu_{\beta})_{V-A}(\overline{c}u)_{V-A}%
\]
For $D^{+}\longrightarrow\pi^{+}\nu_{\alpha}\overline{\nu}_{\beta}$ decay NSIs
is calculated in \cite{C. H. Chen}%

\[
BR(D^{+}\longrightarrow\pi^{+}\nu_{\alpha}\overline{\nu}_{\beta}%
)_{NSI}=|V_{ud}^{\ast}\frac{\alpha_{em}}{4\pi\sin^{2}\theta_{W}}%
\epsilon_{\alpha\beta}^{dL}\ln\frac{\Lambda}{m_{W}}|^{2}BR(D^{+}%
\longrightarrow\pi^{0}e^{+}\nu_{e})
\]
$BR(D^{+}\longrightarrow\pi^{+}\nu_{\alpha}\overline{\nu}_{\beta}%
)_{NSI}=2\times10^{-8}|\epsilon_{\alpha\beta}^{dL}\ln\frac{\Lambda}{m_{W}%
}|^{2}$and it is mentioned that as $\alpha$ and $\beta$ could represent any
lepton, we take $\epsilon_{\tau\tau}^{dL}\sim1,\epsilon_{ll^{/}}^{dL}%
\langle1~$for $l=l^{/}\neq\tau.$ Here $\ln\frac{\Lambda}{m_{W}}\sim1.$

We point out that the same is applicable to two other processes $D_{s}%
^{+}\longrightarrow K^{+}\nu_{\alpha}\overline{\nu}_{\beta}~$and
$D^{0}\rightarrow\pi^{0}\nu_{\alpha}\overline{\nu}_{\beta}.$%

\begin{align*}
BR(D_{s}^{+}  &  \longrightarrow K^{+}\nu_{\alpha}\overline{\nu}_{\beta
})_{NSI}=|V_{ud}^{\ast}\frac{\alpha_{em}}{4\pi\sin^{2}\theta_{W}}%
\epsilon_{\alpha\beta}^{dL}\ln\frac{\Lambda}{m_{W}}|^{2}BR(D_{s}%
^{+}\longrightarrow K^{0}e^{+}\nu_{e})\\
BR(D^{0}  &  \longrightarrow\pi^{0}\nu_{\alpha}\overline{\nu}_{\beta}%
)_{NSI}=|V_{ud}^{\ast}\frac{\alpha_{em}}{4\pi\sin^{2}\theta_{W}}%
\epsilon_{\alpha\beta}^{dL}\ln\frac{\Lambda}{m_{W}}|^{2}BR(\overline{D}%
^{0}\longrightarrow\pi^{-}e^{+}\nu_{e})
\end{align*}
Using PDG 2012 \cite{PGD 2012} Values $BR(D_{s}^{+}\longrightarrow K^{0}%
e^{+}\nu_{e})=(3.7\pm1)\times10^{-3},$ $V_{ud}=0.97425\pm0.00022,$
$\alpha_{em}=\frac{1}{137},$we get%

\[
BR(D_{s}^{+}\longrightarrow K^{+}\nu_{\alpha}\overline{\nu}_{\beta}%
)_{NSI}=2.22796\times10^{-8}(\epsilon_{\alpha\beta}^{dL})^{2}|\ln\frac
{\Lambda}{m_{W}}|^{2}%
\]
For $\epsilon_{\tau\tau}^{dL}\sim1~$and $\ln\frac{\Lambda}{m_{W}}\sim1~,$we
get $BR(D_{s}^{+}\longrightarrow K^{+}\nu_{\alpha}\overline{\nu}_{\beta
})_{NSI}=2.22796\times10^{-8}.$

Similarly for$~BR(\overline{D}^{0}\longrightarrow\pi^{-}e^{+}\nu
_{e})=2.89\times10^{-3}$we have
\[
BR(D^{0}\longrightarrow\pi^{0}\nu_{\alpha}\overline{\nu}_{\beta}%
)_{NSI}=3.21068\times10^{-8}(\epsilon_{\alpha\beta}^{dL})^{2}|\ln\frac
{\Lambda}{m_{W}}|^{2}%
\]
$10^{-8}$will be in the range of BES-III. If it is not detected by there even
than useful limits for new physics can be suggested.%

\begin{tabular}
[c]{|l|l|l|l|l|}\hline
Reaction & SM & NSIs & $\epsilon_{\tau\tau}^{dL}$ &
\begin{tabular}
[c]{l}%
$\epsilon_{ll^{/}}^{dL}$\\
$l=l^{/}\neq\tau$%
\end{tabular}
\\\hline
$BR(D^{+}\longrightarrow\pi^{+}\nu_{\alpha}\overline{\nu}_{\beta})$ &
\begin{tabular}
[c]{|l|l|}\hline
Long Distance & $<8\times10^{-16}$\\\hline
Short Distance & $3.9\times10^{-16}$\\\hline
\end{tabular}
\cite{Slac} & $2\times10^{-8}$ & $\sim1$ & $\langle1$\\\hline
$BR(D_{s}^{+}\longrightarrow K^{+}\nu_{\alpha}\overline{\nu}_{\beta})$ &
\begin{tabular}
[c]{ll}%
Long Distance & $<4\times10^{-16}$\cite{TTP}\\
Short Distance & $1.5\times10^{-16}$%
\end{tabular}
& $2.23\times10^{-8}$ & $\sim1$ & $\langle1$\\\hline
$BR(D^{0}\longrightarrow\pi^{0}\nu_{\alpha}\overline{\nu}_{\beta})$ &
\begin{tabular}
[c]{|l|l|}\hline
Long Distance & $<6\times10^{-16}$\\\hline
Short Distance & $4.9\times10^{-16}$\\\hline
\end{tabular}
\cite{Slac} & $3.21\times10^{-8}$ & $\sim1$ & $\langle1$\\\hline
\end{tabular}
%

\ Table 1.%
%

\begin{tabular}
[c]{l}%
$%
\begin{tabular}
[c]{l}%
$\underset{\text{%
\ Figure 3.%
}~}{%
{\parbox[b]{4.4157in}{\begin{center}
\includegraphics[
natheight=3.947900in,
natwidth=4.364700in,
height=3.9972in,
width=4.4157in
]%
{NEINK202.wmf}%
\\%
\protect\begin{tabular}
[c]{l}%
$Contour~Plot~(D^{0}\longrightarrow\pi^{0}\nu_{\tau}\overline{\nu}_{\tau
})_{NSI}~as~a~function~of~\epsilon_{\tau\tau}^{dL}$\protect\\
$and~new~energy~scale~\Lambda$%
\protect\end{tabular}
$~$
\end{center}}}
}$%
\end{tabular}
\ \ \ \ \ \ \ $%
\end{tabular}

$%
\begin{tabular}
[c]{l}%
$%
\begin{tabular}
[c]{l}%
$\underset{\text{%
\ Figure 4.%
}}{%
{\parbox[b]{3.8821in}{\begin{center}
\includegraphics[
natheight=2.655800in,
natwidth=3.833700in,
height=2.6991in,
width=3.8821in
]%
{NEINK203.wmf}%
\\
$%
\protect\begin{array}
[c]{c}%
~~NSIs~Branching~Ratio~of~D^{0}\rightarrow\pi^{0}\nu_{\tau}\overline{\nu
}_{\tau}\protect\\
\Lambda~is~new~phyics~scale,\epsilon_{\tau\tau}^{dL}~new~physics~parameter
\protect\end{array}
$
\end{center}}}
}$%
\end{tabular}
\ \ \ \ \ \ $%
\end{tabular}
\ \ \ \ \ \ $

\section{NSIs in D$_{s}^{+}\rightarrow D^{+}\nu_{\alpha}\overline{\nu}_{\beta
}$}

It is short distance dominant process represented by quark level process
$s\rightarrow d~\nu_{\alpha}\overline{\nu}_{\beta}~$just like $K^{+}%
\rightarrow\pi^{+}\nu_{\alpha}\overline{\nu}_{\beta}$~for which $\epsilon
_{\tau\tau}^{uL}~\leq\frac{8.8\times10^{-3}}{\ln\frac{\Lambda}{m_{W}}}~$is
pointed out by \cite{C. H. Chen}. NSIs Diagram in fig 3%

\begin{tabular}
[c]{l}%
$\underset{\text{%
\ Figure 5.%
}~}{%
{\parbox[b]{1.5264in}{\begin{center}
\includegraphics[
natheight=1.406200in,
natwidth=1.489200in,
height=1.4416in,
width=1.5264in
]%
{NEINK204.wmf}%
\\
$\ NSIs~s\rightarrow d~\nu_\alpha\overline{\nu}_\beta~$
\end{center}}}
}$%
\end{tabular}
%

\  {}{}{}%

The effective Hamiltonian for such reaction is given by%

\[
H_{eff}^{NSI}=\frac{G_{F}}{\sqrt{2}}(V_{us}^{\ast}V_{ud}\frac{\alpha_{em}%
}{2\pi\sin^{2}\theta_{W}}\epsilon_{\alpha\beta}^{uL}\ln\frac{\Lambda}{m_{W}%
})\times(\nu_{\alpha}\overline{\nu}_{\beta})_{V-A}(\overline{s}d)
\]
From this branching ratio of $D_{s}^{+}\rightarrow D^{+}\nu_{\alpha}%
\overline{\nu}_{\beta}~$for NSIs becomes%

\[
Br(D_{s}^{+}\rightarrow D^{+}\nu_{\alpha}\overline{\nu}_{\beta})_{NSI}%
=|\frac{\alpha_{em}}{4\pi\sin^{2}\theta_{W}}V_{ud}\epsilon_{\alpha\beta}%
^{uL}\ln\frac{\Lambda}{m_{W}}|^{2}BR(D_{s}^{+}\longrightarrow D^{0}e^{+}%
\nu_{e})
\]

Using estimated $BR(D_{s}^{+}\longrightarrow D^{0}e^{+}\nu_{e})=5\times
10^{-6}$~for BES in \cite{Hai-Bo} we get NSIs $Br$($D_{s}^{+}\rightarrow
D^{+}\nu_{\tau}\overline{\nu}_{\tau})=2.33153\times10^{-15}$ which could
enhance SM value $(\sim6\times10^{-15})$~even at electroweak scale.%

\ {}{}{}%
%

\begin{tabular}
[c]{|l|l|l|l|}\hline
$Process$ & $SM$ & $NSIs$ & $\epsilon_{\tau\tau}^{uL}$\\\hline
$D_{s}^{+}\rightarrow D^{+}\upsilon\overline{\upsilon}$ & $6\times10^{-15}$ &
$2\times10^{-15}$ & $O(10^{2})$\\\hline
\end{tabular}
%

\ Table 2%
%

\ {}{}{}%

This can not be detected in BES-III but there is a chance for them in
B-factories or in a future accelerator.The contour plot of Br ratio as a
function of new energy scale $\Lambda$ and $\epsilon_{\tau\tau}^{uL}~$is given
in fig 6.%

\begin{tabular}
[c]{l}%
$\underset{\text{%
\ Figure 6.%
}}{%
{\parbox[b]{4.1018in}{\begin{center}
\includegraphics[
natheight=3.937500in,
natwidth=4.052500in,
height=3.9859in,
width=4.1018in
]%
{NEINK205.wmf}%
\\
$~%
\protect\begin{tabular}
[c]{l}%
$Contour~Plot~(D_{s}^{+}\longrightarrow D^{+}\nu_{\tau}\overline{\nu}_{\tau
})~NSI~as~a~function~of~\epsilon_{\tau\tau}^{uL}$\protect\\
$and~new~energy~scale~\Lambda$%
\protect\end{tabular}
$
\end{center}}}
}$%
\end{tabular}

\begin{tabular}
[c]{l}%
$\underset{\text{%
\ Figure 7.%
}}{%
{\parbox[b]{3.7974in}{\begin{center}
\includegraphics[
natheight=2.395500in,
natwidth=3.749800in,
height=2.437in,
width=3.7974in
]%
{NEINK206.wmf}%
\\
$%
\protect\begin{array}
[c]{c}%
~Branching~Ratio~of~NSI~of~D_{s}^{+}\rightarrow D^{+}\upsilon_{\tau}%
\overline{\upsilon}_{\tau}\protect\\
\Lambda~is~new~phyics~scale,\epsilon_{\tau\tau}^{uL}~is~new~physics~parameter
\protect\end{array}
$
\end{center}}}
}$%
\end{tabular}

\section{Summary and Discussion}

We investigate two processes $D_{s}^{+}\rightarrow K^{+}\upsilon
\overline{\upsilon},$ $D^{0}\rightarrow\pi^{0}\upsilon\overline{\upsilon}$
which are long distance dominated so they are model dependent that is why new
physics could enhance their branching ratio, as it is added into standard
model. It means contribution form NSIs is very large as compared to SM. But,
as for as the short distance (SD) dominated $D_{s}^{+}\rightarrow
D^{+}\upsilon\overline{\upsilon}~$is concerned the margin is only the
difference between SM and experiments. These are calculated by perturbation
theory which is, so far, considered very authentic so we can not discard SM
contribution all altogether in this case. NSIs can improve SD dominated
processes. For example, for short distance dominated process $K^{+}%
\rightarrow\pi^{+}\upsilon\overline{\upsilon}$ the difference between
experiment and theory is $\sim10^{-10}~$so NSIs can not give more contribution
than this. Unfortunately, it is the only experimentally measured semileptonic
process involved two neutrinos in the final state. We are only using the limit
provided by this process in $D_{s}^{+}\rightarrow D^{+}\upsilon\overline
{\upsilon}.$

\section{Conclusion}

We have calculated $D_{s}^{+}\rightarrow K^{+}\upsilon\overline{\upsilon},$
$D^{0}\rightarrow\pi^{0}\upsilon\overline{\upsilon}~$and$~D_{s}^{+}\rightarrow
D^{+}\upsilon\overline{\upsilon}~$decays with branching ratios of
$2.23\times10^{-8},~3.21\times10^{-8}$and $2.33\times10^{-15}$ respectively in
the frame work of NSIs. From these calculations bounds on $\epsilon_{\tau\tau
}^{uL}~$and $\epsilon_{\tau\tau}^{dL}~$are $O(10^{-2})~$and $\sim1$
respectively, $\epsilon_{_{\alpha\beta}}^{dL}<1~$for $\alpha,\beta=e,\mu.$
NSIs are giving much higher values for long distance dominated processes and
there is a considerable enhancement in the short distance processes of $D$
rare decays involving neutrinos in the final state.

\end{document}